\def\Jrnl#1#2#3#4{{#1} {\bf #2}, #3 (#4)}   
\def\EPJB{Eur.\ Phys.\ J.\ {\bf B}}

\def\PLA{Phys.\ Lett.\ \bf A}

\def\PRB{Phys.\ Rev.\ \bf B}   
   
\def\PRL{Phys.\ Rev.\ Lett.\ }    
    
\def\RPP{Rep.\ Prog.\ Phys.\ }

\def\RMP{Rev.\ Mod.\ Phys.\ }    
\def\Section#1{}      
\def\ua{\uparrow}    
\def\da{\downarrow}

\def\beq{\begin{equation}}    
\def\eeq{\end{equation}}    
\def\bea{\begin{eqnarray}}    
\def\eea{\end{eqnarray}}    
\def\nn{\nonumber}

\def\age{\,\raise2pt\hbox{$\mathop{>}\limits_{\raise 2pt    
\hbox{$\sim$}}$}\,}          
\def\ale{\,\raise2pt\hbox{$\mathop{<}\limits_{\raise 2pt    
\hbox{$\sim$}}$}\,}    
\setbox122 = \hbox{1}    
\def\id{\rlap{1}\rlap{\kern 1pt \vbox{\hrule width 4pt depth 0 pt}}    
        \rlap{\kern 4 pt \hbox{\vrule height \ht122 depth 0 pt}}    
           \hskip\wd122}    
\def\Zed{\Bbb Z}   
  
\def\breakon{\end{multicols}\widetext\vspace{-.7cm}    
\noindent\rule{.49\linewidth}{.3mm}\rule{.3mm}{.5cm}\vspace{0.0cm}}    
\def\breakoff{\vspace{-.25cm}    
\noindent    
\rule{.50\linewidth}{.0mm}\rule[-.47cm]{.3mm}{.5cm}  
\rule{.489\linewidth}{.3mm}    
\vspace{-.55cm}    
\begin{multicols}{2}    
\narrowtext\noindent}    
    
\documentstyle[aps,multicol,psfig,amsfonts]{revtex}   
\begin{document}    
%
%
%
%
\title{Hubbard ladders in a magnetic field}    
\author{D.C.\ Cabra$^{1,2}$, A.\ De Martino$^{3}$, P.\ Pujol$^{4}$    
and P.\ Simon$^{5}$}    
\address{    
$^{1}$Departamento de F\'{\i}sica, Universidad Nacional de la Plata,    
	C.C.\ 67, (1900) La Plata, Argentina;\\    
$^{2}$Facultad de Ingenier\'\i a, Universidad Nacional de Lomas de    
	Zamora, Cno. de Cintura y Juan XXIII,\\    
      (1832) Lomas de Zamora, Argentina.\\    
$^{3}$Fakult\"at f\"ur Physik, Albert-Ludwigs-Universit\"at,     
	79104 Freiburg, Germany.\\    
$^{4}$Laboratoire de Physique\cite{URA}, Groupe de Physique Th\'eorique,    
	ENS Lyon,\\    
	 46 All\'ee d'Italie, 69364 Lyon C\'edex 07, France.\\    
$^{5}$Department of Physics and Astronomy, University of British Columbia,    
	V6T 1Z1 Vancouver, B.C., Canada.    
}    
\maketitle    
%
    
\vspace{.3cm}    
    
\begin{abstract}    
The behavior of a two leg Hubbard ladder in the presence of a magnetic field    
is studied by means of Abelian bosonization. We predict the appearance of a   
new (doping dependent) plateau in the magnetization curve of a doped $2-$leg   
spin ladder in a wide range of couplings. We also discuss the extension   
to $N-$leg Hubbard ladders.     
\end{abstract}
\begin{multicols}{2}    
\narrowtext    
    
\vskip -0.2cm    
\vskip2pc    
    
    
In the past few years there has been intense research activity, both   
experimental and theoretical, on ladder systems \cite{DR}. These   
systems interpolate between 1d and 2d and, due to the strong   
quantum fluctuations enhanced by the low dimensionality, exhibit \  
surprising and exotic behaviors, such as, for instance, the so-called   
even-odd effect.

A feature of spin ladders which attracted considerable interest is the    
occurrence of plateaux in their magnetization curves. One of the main   
observations has been the {\em rationality} of the value of the   
magnetization at which plateaux can appear \cite{OYA,Totsuka,CHPspin}.  
However, it has been recently pointed out that plateaux can also  
appear at {\em irrational} values of the magnetization. This happens for  
example in periodically modulated doped Hubbard chains \cite{letter}, but  
this scenario is believed to be generic in doped systems.   
A novel and interesting feature of the plateaux predicted in \cite{letter} is   
the fact that their position depends on the filling and therefore   
can be  moved down to low magnetization values by means of doping.   
This makes doping dependent plateaux potentially observable in   
experiments at low magnetic fields.   
Previous studies revealed doping dependent plateaux in other   
systems, such as the one-dimensional Kondo lattice model \cite{TSU}   
and an integrable spin-$S$ doped  $t-J$ chain \cite{FS}.    
     
The purpose of this letter is to investigate the conditions of occurrence of     
magnetization plateaux in doped ladder systems. We have focussed    
on the case of coupled Hubbard chains, the spin ladders being particular limits   
of Hubbard ladders at half filling.     
Various models of coupled chains have been thoroughly investigated 
in the last decade, but comparatively little attention has been devoted 
to the effects of a magnetic field on their properties.     
In \cite{rice98} (see also \cite{YOA}) it has been  proven that in a $N$-leg   
Hubbard ladder gapless excitations exist at momentum $k=2\pi N n_{\ua,\da}$ if   
the commensurability conditions $N n_{\ua,\da}\in \Zed$ are not satisfied  
\cite{notation}.   
This holds even if the low energy excitations cannot be identified   
as the usual separated charge and spin excitations.    
According to this result, magnetization plateaux at $m\ne 0$   
can only occur if at least one of the conditions   
$N n_{\ua,\da} = {N\over 2} (n \pm m) \in \Zed$   
is satisfied, which considerably restricts the window to find  new plateaux.   
In order to prove the presence of a plateau one has to show   
that there is a finite gap in the total spin sector, since the above requirement  
just provides a necessary condition for the occurrence of a non trivial   
$m\ne 0$ plateau.   
We have found that when the conditions    
\beq     
N n_{\ua,\da}\equiv \frac{N}{2} ( n \pm m ) \in \Zed     
\label{pm}   
\eeq     
are simultaneously satisfied, or when only one of them is satisfied and the   
doping  kept fixed, a plateau can indeed occur for some range of parameters.  
In the last situation, one has a doping dependent plateau which,  
for the specific case of $N=2$, is predicted to be present for a   
wide range of couplings.    
\begin{figure}[h]    
\hspace {1cm} \psfig{figure=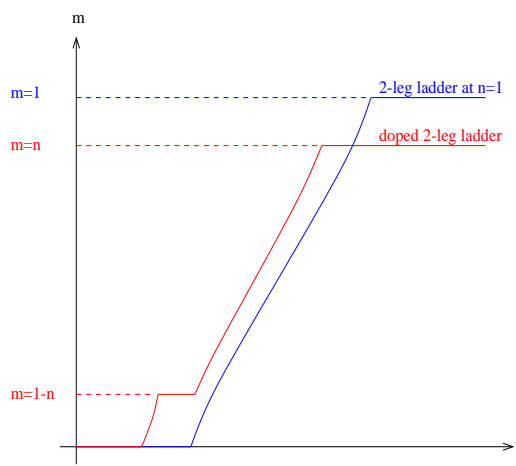,height=6cm,width=7cm}    
\vspace{0.1cm}    
\caption{Schematic magnetization curve of a 2-leg Hubbard    
ladder at $n=1$ (half-filling) and $n\ne 1$. The width of the plateaux   
depends on $t_\perp/t,U/t,n$.  }    
\label{}    
\end{figure}      
Our main result is schematically summarized in Fig.\ 1, which shows   
the expected magnetization curves of a 2-leg Hubbard ladder for both the doped   
and undoped cases. According to our analysis, apart from the known plateau at   
$m=0$, plateaux should also occur at $m=\pm(1-n)$.   
These could be observed experimentally by doping any system described   
by a 2-leg ladder. Possible candidates are notably the spin-${1\over 2}$   
ladder compounds $\mbox{SrCu}_2\mbox{O}_3$ \cite{ishida} and   
Cu$_2$(C$_5$H$_{12}$N$_2$)$_2$Cl$_4$ (which has been studied under high magnetic   
fields \cite{levy}), or some quasi-one-dimensional spin-${1\over 2}$   
organic system. These materials are often difficult to dope.   
Nevertheless, doping has been achieved for the ladder material   
Sr$_{0.4}$Ca$_{13.6}$Cu$_{24}$O$_{41.84}$ under high   
pressure \cite{uehara} (in the latter, superconductivity was also observed).   
Another class of interesting candidates are single wall carbon nanotubes
\cite{nanotube}, whose low energy physics is also described by a model
equivalent to two coupled spin-$1/2$ fermionic chains away 
half filling \cite{egger}. 

Let us turn now to our bosonization analysis (see {\it e.g.} \cite{GNT}).    
We start from the 2-leg Hubbard ladder, with the following lattice   
Hamiltonian:    
\bea    
H = H_I + H_{II} + {\tilde \lambda}_c\sum_{x} n^I_x n^{II}_x + 
{\tilde \lambda}_s   
\sum_{x} {\bf S}_x^{I} {\bf S}_x^{II} \nn \\ + {\tilde \lambda} \sum_x ( c^{I   
\dagger}_{x\uparrow} c^{I \dagger}_{x\downarrow} c^{II}_{x\uparrow}   
c^{II}_{x\downarrow} + c^{II \dagger}_{x\uparrow} c^{II   
\dagger}_{x\downarrow} c^{I}_{x\uparrow} c^{I}_{x\downarrow} )  ~,   
\label{hamgen}    
\eea    
where $H_i$ ($i=I,II$) is the usual Hubbard Hamiltonian:   
\bea    
H_i =   
- t \sum_{x,\alpha} (c^{i   
\dagger}_{x+1,\alpha} c^i_{x,\alpha} + H.c.) + U \sum_x n^i_{x,\uparrow}     
n^i_{x,\downarrow} + \nn \\   
+ \, \mu_i\sum_x ( n^i_{x,\uparrow} + n^i_{x,\downarrow} )     
- {h \over 2} \sum_x ( n^i_{x,\uparrow} - n^i_{x,\downarrow} )   ~,  
\eea  
$c^{i\dagger}_{x,\alpha} , c^{i}_{x,\alpha}$ are electron   
creation and annihilation operators on site $x$ of chain $i$, $\alpha$ is   
the spin index, $ n^i_{x,\alpha}=c^{i\dagger}_{x\alpha} c^i_{x\alpha}$.   
The index $i=I,II$ can be considered either as a chain index or as a band   
index. Accordingly, the Hamiltonian (\ref{hamgen}) corresponds to   
different regimes of coupled chains of interacting electrons.   
These systems have been largely studied analytically,   
for different range of parameters \cite{VZ85}--\cite{LLHR},  
and numerically \cite{DRS,rice,NWS}.   
   
Eq.\ (\ref{hamgen}) describes two identical chains with standard   
Hubbard Hamiltonians $H_{1,2}(t,U_0,\mu_0,h)$ with small 
intrachain repulsion $U_0$, coupled by a direct hopping term:   
\beq     
H_{int} = - t_\perp \sum_{x,\alpha} ( c_{x,\alpha}^{1\dagger}     
c_{x,\alpha}^{2} + H.c. ) ~,    
\eeq     
where $t_\perp$ is the interchain hopping amplitude.    
To see this, one simply has to change variables to the bonding and   
anti-bonding basis:    
\beq    
s_x=\frac{c_x^1 + c^2_x}{\sqrt{2}} ~;~~~    
a_x=\frac{c_x^1 - c^2_x}{\sqrt{2}} ~.   
\eeq     
Rewriting $H_1+H_2+H_{int}$ in this basis and identifying    
the band indices $s$ and $a$ with the indices   
$I$ and $II$, one recovers the Hamiltonian (\ref{hamgen}) with   
$\mu_{I,II} = \mu \mp t_\perp$,   $U = U_0/2$, 
$4 {\tilde \lambda}_c = -{\tilde\lambda}_s =   
-2{\tilde \lambda} = U_0$. Thus we have to deal with two coupled   
inequivalent effective Hubbard chains.    
    
On the other hand, for $U_0 \gg t_\perp$ and close to half filling,    
(namely, for strong repulsive interactions in the individual chains),     
the effect of band splitting is suppressed due to the requirement of 
avoiding double on-site occupancy. In this situation it is more 
appropriate to consider couplings     
directly in terms of the original degrees of freedom of the two individual chains;   
then, $I$ and $II$ have to be identified with the chain index, leading us to two    
equivalent chains coupled by density-density and spin-spin interactions,    
as in (\ref{hamgen}) but with ${\tilde \lambda} =0$ and 
${\tilde \lambda}_{c,s}\sim t_\perp^2/U$.     
    
We are now ready  to treat the two different situations at once, by     
studying the Hamiltonian (\ref{hamgen}), which contains the two    
limits described above. The $m=0$ case has been largely studied in    
the literature \cite{VZ85}-\cite{NWS}. The main conclusions in that case are  
that the spin gap present at half filling survives upon doping, although  
smaller in magnitude, and the appearance of quasi-long-range superconducting   
pairing correlations. The main effect of the magnetic field on the single chain   
is to shift by opposite amounts the up and down Fermi momenta, whose difference is   
proportional to the magnetization, $k_-\equiv k_\ua-k_\da=\pi m$.   
As a  first approximation, we can neglect other effects, such as the mixing   
of spin and charge degrees of freedom, and the dependence of the scaling   
dimensions of the operators on the magnetic field \cite{FK}.     
Note that the calculations can also be handled straightforwardly without   
these approximations, but equations become cumbersome   
(see for example \cite{letter}).  
  
In order to write the bosonized expressions for the two cases for $m\ne 0$   
we have to define the Fermi momenta $k^{I,II}_{\ua,\da}$ separately for each   
band (or chain) and linearize the dispersion relations around them.   
We assume that interactions do not change significantly the free Fermi momenta.   
In the process of bosonization we obtain operators containing  
oscillating factors made of various combinations of the $k^i_\alpha$.    
In the case of inequivalent chains, most of them will be commensurate only if  
specific conditions depending on $t_\perp$ are satisfied. We will neglect in  
the following these non-generic situations and will take into account only  
operators that can be commensurate independently of the value of $t_\perp$.   
Moreover, the operator responsible for the $m=0$ plateau is only commensurate   
for that value of $m$ and hence it will not enter in the following   
discussion.   
  
\vspace{.3cm}   
   
\noindent {\it i) Equivalent chains}     
    
\vspace{.3cm}   
    
In this case $k^I_\alpha = k^{II}_\alpha$, $I,II$ represent chain indices  
and we use the notation $I\equiv 1,II\equiv 2$.  
The continuum Hamiltonians $H_i$ are written as    
\beq    
H_i=\sum_{\nu=c,s}  \frac{v_\nu}{2} \int dx     
[K_\nu (\partial_x \Theta^i_\nu)^2 + K_\nu^{-1}(\partial_x \Phi^i_\nu)^2] ~,    
\label{echam}  
\eeq    
where $v_{c,s}, K_{c,s}$ are the velocities and the effective Luttinger    
parameters for charge and spin degrees of freedom.    
Away from half-filling ($n\neq 1$) and for non zero magnetization \cite{conv}),     
the effective interaction Hamiltonian reads    
\bea    
{\cal H}_{int} = && \lambda_c  \, \partial_x \Phi^1_c \partial_x \Phi^2_c +      
{\lambda_s\over 4} \, \partial_x \Phi^1_s \partial_x \Phi^2_s + \nn \\   
&& -  
\, \lambda_1  \left( - 
\cos [ \sqrt{4\pi} \Phi^-_c ] \cos [ \sqrt{4\pi} \Phi^-_s ] \, + \right. \nn \\    
&& + \, \cos [ 2 (k^1_\da + k^2_\da ) x - \sqrt{8\pi} \Phi^+_\da ] + \nn \\    
&& + \, \left. 
\cos [ 2 ( k^1_\ua + k^2_\ua ) x - \sqrt{8\pi} \Phi^+_\ua ] \, \right)  + \nn \\    
&& + \, \lambda_2 \,     
\cos[ \sqrt{4\pi} \Phi^-_c ]  \cos [ \sqrt{4\pi} \Theta^-_s] ~,    
\label{12Ham}    
\eea    
where   
$\Phi_{c/s}^{i}={1\over\sqrt{2}}(\Phi_\ua^{i}\pm\Phi_\da^{i})$,   
and $\Phi_{}^{\pm}={1\over\sqrt{2}}(\Phi^{1}\pm\Phi^{2})$ \cite{nota}.   
We have kept only the more relevant operators for the case of large $U$.  
     
The bosonic bilinear terms can be absorbed in the kinetic part of the   
Hamiltonian by moving to the $\pm$ basis defined above.   
As a consequence, the velocities and more importantly the effective Luttinger   
parameters are renormalized in the following way:      
\bea   
&&K_{c} \rightarrow K^{\pm}_{c} = K_{c} (1 \pm \lambda_{c}   
K_{c}/v_{c})^{-1/2} ~,\label{renKc}\\  
&&K_{s} \rightarrow K^{\pm}_{s} = K_{s} (1 \pm \lambda_{s}   
K_{s}/4v_{s})^{-1/2} ~.    
\label{renKs}   
\eea    
{}From the above equations we see that in the   
charge sector, the Luttinger parameter for the total charge field   
(symmetric combination)  is slightly reduced with respect to $K_c$ whereas   
the one for the relative charge (antisymmetric combination) is slightly   
increased  (we assume $\lambda_c > 0$).   
Let us recall that $K_c$ decreases from $1$ for $U=0$ to $1/2$ for   
infinite repulsion, thus $K^+_c < 1$.     
We can repeat the analysis above for the spin sector and the  
scaling dimensions of the perturbing operators in eq.\ (\ref{12Ham}) at zero  
loop are then given by:     
\bea    
\cos [ \sqrt{4\pi} \Phi^-_c ] \cos [ \sqrt{4\pi} ~   
\Phi^-_s ] &\rightarrow & K^-_c + K^-_s ~, \\    
\cos [\sqrt{8\pi}\Phi^+_{\da,\ua}] &\rightarrow & K^+_c + K^+_s ~, \\   
\cos [ \sqrt{4\pi}\Phi^-_c ] \cos [ \sqrt{4\pi} \Theta^-_s ]    
&\rightarrow & K^-_c + 1/K^-_s ~.   
\eea    
Due to the double cosine terms in eq.\ (\ref{12Ham}) we conclude 
that both spin and charge antisymmetric sectors are massive. 
Let us now consider the symmetric sectors, which are affected by 
the terms in the third and fourth lines of eq.\ (\ref{12Ham}). 
These operators are commensurate only if conditions (\ref{pm}) are  
satisfied. These appear to be exactly the same conditions  we found 
in the dimerized Hubbard chain for the appearance of plateaux \cite{letter}.    
When the two conditions are simultaneously satisfied we expect both   
$\Phi_s^+$ and $\Phi_c^+$ to be massive.    
Indeed, both $K^+_c$ and $K^+_s$ are decreased with respect to their values   
in the absence of interchain coupling, and we can conclude that the operators,   
when commensurate, are relevant \cite{note1}.    
When instead only one of these conditions is satisfied,   
let us say the one with the $+$ sign, only the field $\Phi^+_\ua$  
gets a relevant cosine interaction. Then following the analysis in  
\cite{letter} we can conclude that a magnetization plateau occurs   
at $m=1-n$ if the total density is kept fixed.  
    
\vspace{.3cm}   
   
\noindent {\it ii) Inequivalent chains}   
   
\vspace{.3cm}   
In this case, $I,II$ are band indices. When $t_\perp > 2t$,   
the bands are splitted and the situation is completely analogous   
to one found in the doped $p$-merized Hubbard chain we treated  in  
\cite{letter}. Following the same lines,   
we can prove that a plateau occurs when $2 n_{\ua,\da} \in \Zed$.   
Notice that in this particular regime the plateau at $m=0$ is no 
longer present. When $t_\perp \sim  t$, non trivial processes between 
bands are allowed and in this case the effective interaction Hamiltonian 
away from half-filling (and at $m\ne 0$) reads     
\bea    
{\cal H}_{int} =    
&&\lambda_c \, \partial_x \Phi^I_c \partial_x \Phi^{II}_c +    
{\lambda_s\over 4} \, \partial_x \Phi^I_s \partial_x \Phi^{II}_s  + \nn \\   
&& - \, \lambda_1
\left( \, \cos [ 2 ( k^I_\da + k^{II}_\da ) x - \sqrt{8\pi} \Phi^+_\da ]   
\, + \right. \nn \\   
&& + \, \left.    
\cos [ 2 ( k^I_\ua + k^{II}_\ua ) x - \sqrt{8\pi} \Phi^+_\ua ] \,   
\right)  + \nn \\    
&& + \, \lambda_3 \, \cos[ \sqrt{4\pi} \Theta_s^- ]     
\cos[\sqrt{4\pi} \Theta_c^- ] ~.    
\label{saHam}    
\eea
In this expression, we have neglected all terms containing combinations of   
$k_\alpha^i$ depending explicitly on $t_\perp$.   
The last marginal term in eq.\ (\ref{saHam}) is generated radiatively.  
The derivative terms lead to a renormalization of the Luttinger   
parameters as in eqs.\ (\ref{renKc}-\ref{renKs}).   
      
As in the strong $U$ limit, when $2n_{\ua,\da}\in \Zed$, the second and   
third line of (\ref{saHam}) become commensurate. We can now repeat the   
analysis concerning the relevance of these operators. Their dimensions    
are determined by the forward scattering terms which can   
be computed at first order in $U\ll t,t_\perp$. It is straightforward 
to see that   
$K_{c,s}^+$ is renormalized from 1  
to $K_{c,s}^+ < 1$, which   
makes the term $\cos[ \sqrt{8\pi} \Phi^+_\ua ]$ (respectively $\cos[   
\sqrt{8\pi} \Phi^+_\da ]$) marginally relevant.   
{}From this and the previous analysis, we can therefore conclude that   
both in weak and in strong coupling regime  
of the 2-leg Hubbard ladder, a magnetization plateau is expected   
when $2n_{\ua,\da}\in \Zed$, and the magnetization curve at density $n<1$   
should have the schematic form depicted in Fig.\ 1.   
  
\vspace{.2cm}   

It is interesting to analyze the behaviour of the diagonal components 
of the charge density wave (CDW) and superconducting pairing (SC) 
operators at a doping dependent plateau, as it occurs {\it e.g.} 
for $m=1-n$.  
In the case of equivalent chains the behaviour of the correlators of   
such order parameters depends on which operator, between the first and   
the last double cosine terms in eq.\ (\ref{12Ham}) dominates.   
When the first term dominates, one could find among diagonal SC and CDW  
operators \cite{operators}  
\bea  
{\cal O}_{\mbox{\tiny CDW},\downarrow}^\pm & \sim &   
e^{i\sqrt{2\pi}\Phi_\downarrow^+}    
\left(  
e^{i\sqrt{2\pi}\Phi_\downarrow^-} \pm e^{-i\sqrt{2\pi}\Phi_\downarrow^-}  
\right) ~,\nn \\  
{\cal O}_{\mbox{\tiny SC}\downarrow \downarrow}^\pm & \sim &   
e^{i\sqrt{2\pi}\Theta_\downarrow^+}   
\left(  
e^{i\sqrt{2\pi}\Phi_\downarrow^-} \mp e^{-i\sqrt{2\pi}\Phi_\downarrow^-}  
\right) ~,  
\label{sc}  
\eea  
algebraic decaying correlators. On the contrary, when it is the last one which  
dominates, then all correlators decay exponentially. One can argue that   
varying the values of $\lambda_s$ and the magnetization, both situations   
could be achieved, but this requires a more detailed analysis of the Bethe 
Ansatz  equations in the presence of a magnetic field.  
For the case of inequivalent chains the analysis is similar but more   
involved, and will  be discussed elsewhere \cite{nleg}. 
However, we find also in this case indications for quasi-long-range order 
of the pairing order  parameter.   
    
The generalization of the above analysis to $N$-leg Hubbard ladders can be   
performed along the same lines, and the presence of a non-trivial plateau when   
$N  n_{\ua,\da}\equiv n \pm m \in \Zed $ could be predicted. It is important to   
stress however that for the operator responsible for such a plateau to be   
relevant a fair amount of interchain coupling would be needed.  
This situation is reminiscent of what was found in the study of $N-$leg spin  
ladders, where the interchain coupling should be strong enough for the  
appearance of non-trivial plateaux \cite{CHPspin}. Details of this analysis  
will be presented elsewhere \cite{nleg}.      
    
    
The behavior of a 2-leg doped Hubbard ladder in the presence of a   
magnetic field generalizes the 2-leg spin ladder case.   
Our main result is that the magnetization curve should present two   
plateaux for a non-trivial value of the magnetization (see Fig.\ 1).    
The plateau at zero magnetization persists for non zero doping,  
while the doping dependent one can be obtained only by keeping the doping of   
the system fixed (as it is the case in most experimental settings).   
This should be contrasted to the dimerized Hubbard chain, in which   
the zero magnetization plateau at half filling is shifted to a non-trivial   
value when the system is doped.   
More importantly, the doping dependent plateaux are predicted for a wide range   
of couplings, which should allow for their experimental observation at quite low   
magnetic fields.   
   
\vskip 0.15cm   
    
    
We acknowledge interesting discussions with A. Honecker and K. Le Hur.  
This work has been partially supported by the EC TMR Programme   
{\it Integrability, non perturbative effects and symmetry in quantum field theory}   
(A.D.M. and P.S.).  
The research of D.C.C. is partially supported by CONICET and   
Fundaci\'on Antorchas, Argentina (grant No.\ A-13622/1-106). 
A.D.M. was also supported in part by the DFG under the Gerhard-Hess 
program and P.S. by NSERC of Canada.    
    

\end{multicols}    
   

\begin{references}
\bibitem{notation} {}$n_{\ua,\da}$ are the densities of spin-up (-down)  
electrons. The total density is $n=n_\ua + n_\da$ (half filling is $n=1$)   
and the magnetization is $m=n_\ua -n_\da$.
\bibitem{conv} Our conventions are the following:  $2 \pi n =   
\sum_{i,\alpha} k^i_\alpha$, $2 \pi m= \sum_{i,\alpha} \alpha k^i_\alpha$.   
For equivalent chains $k^I_\alpha= k^{II}_\alpha \equiv k_\alpha$ and one   
obtains $ \pi n = \sum_{\alpha} k_\alpha\equiv k_+$ and $ \pi m =   
\sum_{\alpha} \alpha k_\alpha\equiv k_-$.
\bibitem{nota} Strictly speaking for $h \ne 0$ the fields $\Phi_{c,s}$   
do not diagonalize the effective Hubbard Hamiltonian, thus eq. (\ref{echam})  
is an approximation.
\bibitem{note1} The relevance of these operators can be established 
explicitly for some particular values of the couplings $U$, $t_{\perp}$ 
and the magnetization $m$ by solving numerically the Bethe ansatz 
equations \cite{FK}.
\end{references}
\end{document}